\newcommand{\mathrm}[1]{\hbox{\rm #1}}

\def\bold#1{\setbox0=\hbox{$#1$}%
      \kern-.02em\copy0\kern-\wd0
      \kern.04em\copy0\kern-\wd0
      \kern-.02em\raise.0433em\box0 }

\def\bdsmall#1{\setbox0=\hbox{$#1$}%
      \kern-.015em\copy0\kern-\wd0
      \kern.03em\copy0\kern-\wd0
      \kern-.015em\raise.0233em\box0 }

\documentstyle[12pt]{article}
\begin{document}
\rm
\centerline{\bf{THE (e,e$'$p) REACTION AND THE}}
\centerline{\bf{TRANSITION TO THE EIKONAL REGIME}}

\vskip1.2truecm
\centerline{\small{A.~Bianconi and M.~Radici}}

\centerline{\small \it{ Dipartimento di Fisica Nucleare e Teorica,
Universit\`a di Pavia, and}}
\centerline{\small \it{ Istituto Nazionale di Fisica Nucleare,
Sezione di Pavia,}}
\centerline{\small \it{ v. Bassi 6, 27100 Pavia, Italy}}

\vskip2.truecm

\begin{abstract}

The reliability of the models describing the Final-State Interactions (FSI)
in (e,e$'$p) scattering at high proton energies is an important issue in
view of the experiments planned at CEBAF. One of the most
popular approaches adopted, the Glauber method, involves the linearization
of the wave equation for the ejected proton travelling through the residual
nucleus. We have studied the consequences of such an assumption for the case
of the $^{12}\mathrm{C(e,e}'\mathrm{p)}^{11}\hbox{\rm B}^*$ reaction at high
proton momenta by comparing the results with the predictions obtained when the
second-order differential equation for the proton wave is solved exactly
for each partial wave. We find that the two methods give
well-correlated angular distributions for momenta in the range 1-4 GeV/c,
i.e. for kinematics relevant to the transition to the eikonal regime.

\end{abstract}

\vskip 1.5cm

In the experiments planned at CEBAF concerning (e,e$'$p)
scattering~\cite{CEBAF}, where
the proton momentum can be larger than 1 GeV/c, a central issue is
the reliability of models describing the Final-State Interactions (FSI)
between the outgoing proton and the residual nucleus. For example, at
moderate missing momenta $p_{\mathrm{m}}$ (${\bold p}_{\mathrm{m}} =
{\bold p} - {\bold q}$, where ${\bold p}$ is the outgoing
proton momentum and $\bold q$ is the momentum transferred to the target by
the electron) an accuracy within $10\%$ is required to unambigously identify
exotic effects like Colour Transparency~\cite{CT}, if any.

The most widely used approach to the problem of FSI at high energy is the
Glauber approximation~\cite{Glauber}, since there is a long
well-established tradition in the analysis of proton-nucleus elastic
scattering~\cite{pp,pprev}. In particular, in that
context it has been shown that the validity of such an approach can arise
from a non trivial cancellation among the leading corrections to the
lowest-order theory~\cite{wall}. However, the generalization to the (e,e$'$p)
scattering is not straightforward because the initial proton state and, in
general, the kinematics are quite different.

In the Impulse Approximation, the basic ingredient of completely exclusive
(e,e$'$p) reactions is the scattering amplitude~\cite{frumou,bgprep}

\begin{equation}
J^{\mu}_{\alpha} ({\bold q}) = \int \hbox{\rm d} {\bold r} \hbox{\rm d}
\sigma \hbox{\rm e}^{{\scriptstyle \mathrm{i}} {\bdsmall {\scriptstyle q}}
\cdot {\bdsmall {\scriptstyle r}}} \chi^{\left( -\right)\, *} ({\bold r},
\sigma) {\hat J}^{\mu} ({\bold q}, {\bold r}, \sigma) \Psi_{\alpha}
({\bold r}, \sigma) \quad ,\label{eq:scattampl}
\end{equation}
where $\chi^{\left( - \right)}$ and $\Psi_{\alpha}$ describe the scattering-
and bound-state wave
functions of the nucleon knocked out from a hole with quantum numbers
$\alpha$, respectively. Usually, the current operator ${\hat J}^{\mu}$ is
approximated by a nonrelativistic expansion in powers of the inverse nucleon
mass, thus introducing uncertainties which become more important with
increasing energy~\cite{vanord,gapieka}. However, our interest
is not focussed on the comparison
with experimental data, but on the analysis of the scattering wave
$\chi^{\left( -\right)}$. Therefore, we have considered the simplified
picture where we retain just the longitudinal component of ${\hat J}^{\mu}$
in the leading order $o(1)$ of the nonrelativistic approximation and we
neglect the nucleon form factor. The cross section becomes, therefore,
proportional to

\begin{equation}
\Big \vert \int \hbox{\rm d} {\bold r} \hbox{\rm d} \sigma \hbox{\rm e}^{
{\scriptstyle \mathrm{i}} {\bdsmall { \scriptstyle q}} \cdot {\bdsmall
{\scriptstyle r}} } \chi^{\left( - \right) \, *} ({\bold r}, \sigma)
\Psi_{\alpha} ({\bold r}, \sigma) \Big \vert^2 \equiv
S^{\mathrm{D}}_{\alpha} ({\bold q}) \quad ,
\label{eq:specdist}
\end{equation}
which is traditionally identified as the ``distorted'' spectral density
$S^{\mathrm{D}}_{\alpha}$~\cite{bgp82} at the missing energy corresponding
to the knockout hole $\alpha$.

In the framework of the Distorted-Wave Impulse Approximation
(DWIA) \cite{frumou,bgprep} the
scattering wave function $\chi^{\left( - \right)}$ is solution of
the Schr\"odinger equation

\begin{equation}
\left( - {\hbar^2 \over {2 m}} \nabla^2 + V \right) \chi = E_{
\mathrm{cm}} \chi \quad ,
\label{eq:schroeq}
\end{equation}
where $m$ is the reduced mass of the proton in interaction with the residual
nucleus, $E_{\mathrm{cm}}$ is its kinetic energy in the cm system and $V$
contains a local equivalent energy-dependent optical potential effectively
describing the residual interaction.

Eq. (\ref{eq:schroeq}) can be solved for each partial wave of $\chi^{\left( -
\right)}$ up to a maximum angular momentum $L_{\mathrm{max}} (p)$, where
a convergency criterion for the partial-wave expansion is satisfied. The
boundary condition is such that each incoming partial wave
coincides asymptotically with the corresponding component of the plane wave
associated to the proton momentum ${\bold p}$. Typically, this method (from
now on method I) has been applied to (e,e$'$p) scattering for proton momenta
below 0.5 GeV/c and $L_{\mathrm{max}} < 50$~\cite{bgprep}. In the kinematics
explored in this work a maximum $L_{\mathrm{max}} = 120$ has been used.

At higher energies the Glauber method~\cite{Glauber} suggests an alternative
way (from now on method II) of solving eq. (\ref{eq:schroeq}) by linearizing
it along the propagation axis $\hat z$:

\begin{eqnarray}
{\bold r} &\equiv& z {\displaystyle {{\bold p} \over p}} + {\bold b}
\label{eq:zb} \\
\nabla^2 &\simeq& {\displaystyle {\partial^2 \over {\partial z^2}}}
\label{eq:nabla} \\
\left( {\partial^2 \over {\partial z^2}} + p^2 \right) &=& {\displaystyle
\left( {\partial \over {\partial z}} + \hbox{\rm i} p \right) \cdot \left(
{\partial \over {\partial z}} - \hbox{\rm i} p \right)} \nonumber \\
&\simeq& {\displaystyle 2 \hbox{\rm i} p \cdot \left( {\partial \over
{\partial z}} - \hbox{\rm i} p \right)} \quad ,
\label{eq:glau}
\end{eqnarray}
where ${\bold b}$ describes the degrees of freedom transverse to the motion
of the struck particle with momentum ${\bold p}$. With this approximation eq.
(\ref{eq:schroeq}) becomes

\begin{equation}
\left( {\partial \over{\partial z}} - \hbox{\rm i} p \right) \chi = {1 \over
{2 \mathrm{i} p}} \, V \chi \quad .
\label{eq:schroglau}
\end{equation}
The boundary condition is of incoming unitary flux of plane waves.

Both methods I and II solve the Schr\"odinger equation for the nucleon
scattering wave. Relativistic effects are correctly taken into account
only in a proper calculation of the kinematics. In the case of the
application of the Glauber approach to unpolarized proton-nucleus scattering,
this approximation does not seem to produce relevant
consequences~\cite{pp,pprev,wall}.
Nevertheless, because of the previously mentioned differences, the
generalization to the (e,e$'$p) should be tested~\cite{vanord,amapieka,rost}.

The eikonal approximation is supposed to reproduce the
exact solution of eq. (\ref{eq:schroeq}) for $p_{\mathrm{m}} \ll q$, and in
general its reliability increases with the ejectile energy~\cite{Glauber},
ideally in the limit
where $\chi^{\left( - \right)}$ is expanded on an infinite number of partial
waves. On the other hand, method I can be
considered reliable only for nucleon energies such that the condition
$L_{\mathrm{max}} \gg R_{\mathrm{target}} \, p$ is fulfilled.

In a previous work~\cite{noi} we have analysed the spectral density
$S^{\mathrm{D}}_{\alpha}$ of eq. (\ref{eq:specdist}) for the
$^{12}\mathrm{C(e,e}'\mathrm{p)}^{11}\mathrm{B}_{\mathrm{s} 1/2}$ reaction
in parallel kinematics (${\bold p}$ along ${\bold q}$). In the range
$1 < p < 2$ GeV/c and for $L_{\mathrm{max}} = 120$ we found a good
correlation between the predictions of methods I and II, the (small)
discrepancies being ascribed to the impossibility of taking into account in
eq. (\ref{eq:schroglau}) the interference between the incoming and the
reflected flux. The consequent overestimation of $S^{\mathrm{D}}_{\alpha}$
with respect to method I was found to be related to kinematics and
proportional to the absorption strength of the optical potential.

In the present work we have extended the calculations up to $p,q = 4$ GeV/c
by improving the numerical precision of the {\tt FORTRAN} code. We have
considered
the $^{12}\mathrm{C(e,e}'\mathrm{p)}^{11}\mathrm{B}^*$ reaction in the
socalled perpendicular kinematics, where $p$ and $q$ are kept constant and
the angle between their directions, $\gamma$, is allowed to vary. The bound
state
$\Psi_{\alpha}$ in eq. (\ref{eq:specdist}) is a solution of the Woods-Saxon
potential of Comfort and Karp~\cite{bound} with the quantum numbers $\alpha$
of the s wave. For sake of simplicity, the contribution to $V$ coming from the
Coulomb potential has been neglected to avoid numerical problems related to
the high angular momenta required. Therefore, in proper terms the results
presented here refer to the (e,e$'$n) reaction. $V$ is an optical potential
of the simple form

\begin{eqnarray}
V(r) &=& \left( U + \hbox{\rm i} W \right) \, {\displaystyle {1 \over {1 +
\hbox{\rm e}^{{{r - R} \over a}}}}} \nonumber \\
&\equiv& \left( U + \hbox{\rm i} W \right) \, \rho (r) \quad ,
\label{eq:opt}
\end{eqnarray}
with $R = 1.2 \times A^{1/3}$ fm and $a = 0.5$ fm. The nuclear
density $\rho (r)$ defined in eq. (\ref{eq:opt}) is normalized such that
$\rho (0) = 1$.

At the nucleon energies here considered, the parameters $U,W$ can only be
guessed. According to the Glauber model the imaginary part should scale as
$W \sim p / 10$ MeV, while $U / W$ should equal the ratio between the real
and the imaginary parts of the average proton-nucleon forward-scattering
amplitude, which is expected to be small in the considered
kinematics~\cite{lech}.

In fig.1 $S^{\mathrm{D}}_{\mathrm{s} 1/2}$ is calculated by method I
as a function of $V$ for $p_{\mathrm{m}} = 0$ and $p = q = 1.4$ GeV/c, where
the elementary cross section for the rescattering of the ejectile is
predominantly inelastic. The three curves
correspond to $W = 0, \, 50$ and $100$ MeV, for $U$ ranging continuously from
$-50$ to $+50$ MeV. The $(U=0, \, W=0)$ point corresponds to the Plane-Wave
Impulse Approximation (PWIA) result, where any rescattering between the
ejectile and the residual nucleus is neglected. The middle curve ($W =
50$ MeV) shows an average $40 \%$ damping with respect to the PWIA result
in agreement with the observation of the NE18 experiment in the context of a
semi-inclusive (e,e$'$p) reaction at small $p_{\mathrm{m}}$~\cite{NE18}.
It is evident
that the sensitivity to both the sign and the magnitude of the real part of
the potential is very small but for huge values of $U \gg W$, which are
forbidden by the mainly absorbitive character of the proton-nucleon
amplitude at these kinematics. Therefore, in this work we will use $U$ = 0.

Eq. (\ref{eq:schroglau}) produces by definition
results that only depend on $V$ through the ratio $V/p$. If $V$ is a linear
function of $p$ (as suggested by the Glauber model), a constant absorption
is produced at any value of momentum transfer, provided that $p = q$ and
$p_{\mathrm{m}} \simeq 0$ (i.e. small $\gamma$). In fig. 2 we show that the
same property holds, with a good approximation, also for method I. The
distorted spectral density is given as a function of $p = q$ for
$p_{\mathrm{m}} = 0$ and the choice $U=0, \, W=p \, 50/1400$ MeV, which
produces at $p = q = 1.4$ GeV/c the $40 \%$ absorption observed in the NE18
experiment. The flat curve shows that this damping remains constant down to
very low values of $p = q$ ($p = q \simeq 0.4$ GeV/c). However, the choice of
an absorption-dominated potential is reasonable only in the kinematical
region where the elementary proton-nucleon scattering amplitude is dominated
by inelastic processes, i.e. for $p \ge 1$ GeV/c~\cite{lech}.
Only above this threshold, which is anyway relevant to the kinematics
explored at CEBAF, our considerations can be applied (for applications
of the Glauber approach to (e,e$'$p) reactions at lower energies see
refs.~\cite{pandha,marco}). Since the NE18 observation of a roughly
$q$-independent $40 \%$ damping at small $p_{\mathrm{m}}$ is reproduced
by both methods I and II with the optical potential of eq. (\ref{eq:opt}) and
with $U=0, \, W=p \, 50/1400$ MeV, we adopt this choice in the following also
for the calculation of the angular distributions.

In the pure Glauber theory $W$ is not a free parameter and would
result in a larger value. As previously mentioned, the increase of $W$
leads to a larger discrepancy between the results of methods I and II,
but the overall agreement is not too much spoiled~\cite{noi}.
However, the $W$-value suggested by the Glauber model produces a sensibly
larger absorption than observed in the NE18 experiment (for more details
on this topic see refs.~\cite{pandha,niko,jap,ffstrik}).

In figs. 3 and 4 the distorted spectral density
$S^{\mathrm{D}}_{\mathrm{s} 1/2}$ is shown as a function of $\gamma$ for
$q = p = 1.4$ and $4$ GeV/c, respectively. Because of the large range of
angles considered, several diffraction minima are explored while the size of
the distribution falls down by many orders of magnitude. The dotted line
represents the result with no final-state interactions (PWIA), which is of
course identical in both methods. The solid and the dashed lines are the
results of method I and II, respectively. The two angular distributions are
rather well correlated in all the kinematics here explored, except in the
diffraction minima for $p = q = 1.4$ GeV/c. We have checked that these
discrepancies are smoothed with increasing $p,q$ until they almost disappear at
$p = q = 4$ GeV/c, as it is clear in fig. 4. In any case, through all the
kinematics considered the oscillatory patterns
are very close to each other across a remarkably large range of
variation in size.

It must be noticed that the rich diffractive pattern of the angular
distributions is partially due to the nontrivial structure of the PWIA
contribution, which itself contains many local minima. We have already shown
in a previous work~\cite{noi} that if the Woods-Saxon bound state
$\Psi_{\alpha}$ is
substituted by a pure harmonic oscillator, so to produce an exponentially
decreasing angular distribution in PWIA, the results of both methods I and II
still show an oscillatory pattern at large angles due to FSI. Thus, the
natural interpretation is that the diffractive minima, which are reminiscent
of the angular distribution for elastic proton-nucleus scattering~\cite{pp},
derive
from the fact that the ejected proton is testing coherently the residual
nucleus. This is peculiar of a completely exclusive reaction, where the
residual nucleus does not fragment. Energy-integrated
distributions (i.e. for a semi-inclusive (e,e$'$p)
reaction~\cite{niko,andrea}) are by definition less sensitive
to the structure of the recoiling ($A-1$) system, thus leading to very
different angular shapes.

\vskip .5cm

We have shown that for the
$^{12}\mathrm{C(e,e}'\mathrm{p)}^{11}\mathrm{B}_{\mathrm{s} 1/2}$
reaction and for outgoing proton momenta in the range $1 < p < 4$ GeV/c
(relevant to the planned experiments at CEBAF) the eikonal approximation to
the scattering wave of the ejectile produces FSI effects very similar to the
ones obtained when the complete second-order differential equation is solved
exactly up to $120$ partial waves. The angular distributions are in good
agreement up to very large angles, where the absolute size can fall down by
many orders of magnitude. The observed oscillatory pattern can be interpreted
as a coherent diffractive scattering between the ejectile and the residual
nucleus. Therefore, completely exclusive (e,e$'$p) reactions are best suited
to verify this prediction.

\vskip .5cm

We would like to thank O. Benhar, S. Boffi, S. Jeschonnek, N.N. Nikolaev and
S. Simula for many stimulating discussions.


\newpage


\centerline{Captions}

\vspace{2cm}

\begin{itemize}

\item[Fig. 1 - ] The distorted spectral density
$S^{\mathrm{D}}_{\mathrm{s} 1/2}$ as a function of the depth $U$ of the real
part of the optical potential for the
$^{12}\mathrm{C(e,e}'\mathrm{p)}^{11}\mathrm{B}_{\mathrm{s} 1/2}$ reaction
at the outgoing proton momentum $p = 1.4$ GeV/c and at the missing momentum
$p_{\mathrm{m}} = 0$. The dashed line corresponds to the depth $W = 0$ of the
imaginary potential, the solid line to $W = 50$ MeV, the dotted line to
$W = 100$ MeV.

\end{itemize}

\vspace{1cm}

\begin{itemize}

\item[Fig. 2 - ]  The distorted spectral density
$S^{\mathrm{D}}_{\mathrm{s} 1/2}$ as a function of $p = q$ for the
$^{12}\mathrm{C(e,e}'\mathrm{p)}^{11}\mathrm{B}_{\mathrm{s} 1/2}$ reaction
at $p_{\mathrm{m}} = 0$ and with the optical potential depths
$U=0, \, W=p \, 50/1400$ MeV.

\end{itemize}

\vspace{1cm}

\begin{itemize}

\item[Fig. 3 - ] The distorted spectral density
$S^{\mathrm{D}}_{\mathrm{s} 1/2}$ as a function of the angle $\gamma$ between
${\bold p}$ and ${\bold q}$ for the
$^{12}\mathrm{C(e,e}'\mathrm{p)}^{11}\mathrm{B}_{\mathrm{s} 1/2}$ reaction
with $p = q = 1.4$ GeV/c. The dotted line shows the PWIA result. The solid
and dashed lines are the results of methods I and II, respectively (see text).

\end{itemize}

\vspace{1cm}

\begin{itemize}

\item[Fig. 4 - ] The same as in fig. 3, but for $p = q = 4$ GeV/c.

\end{itemize}

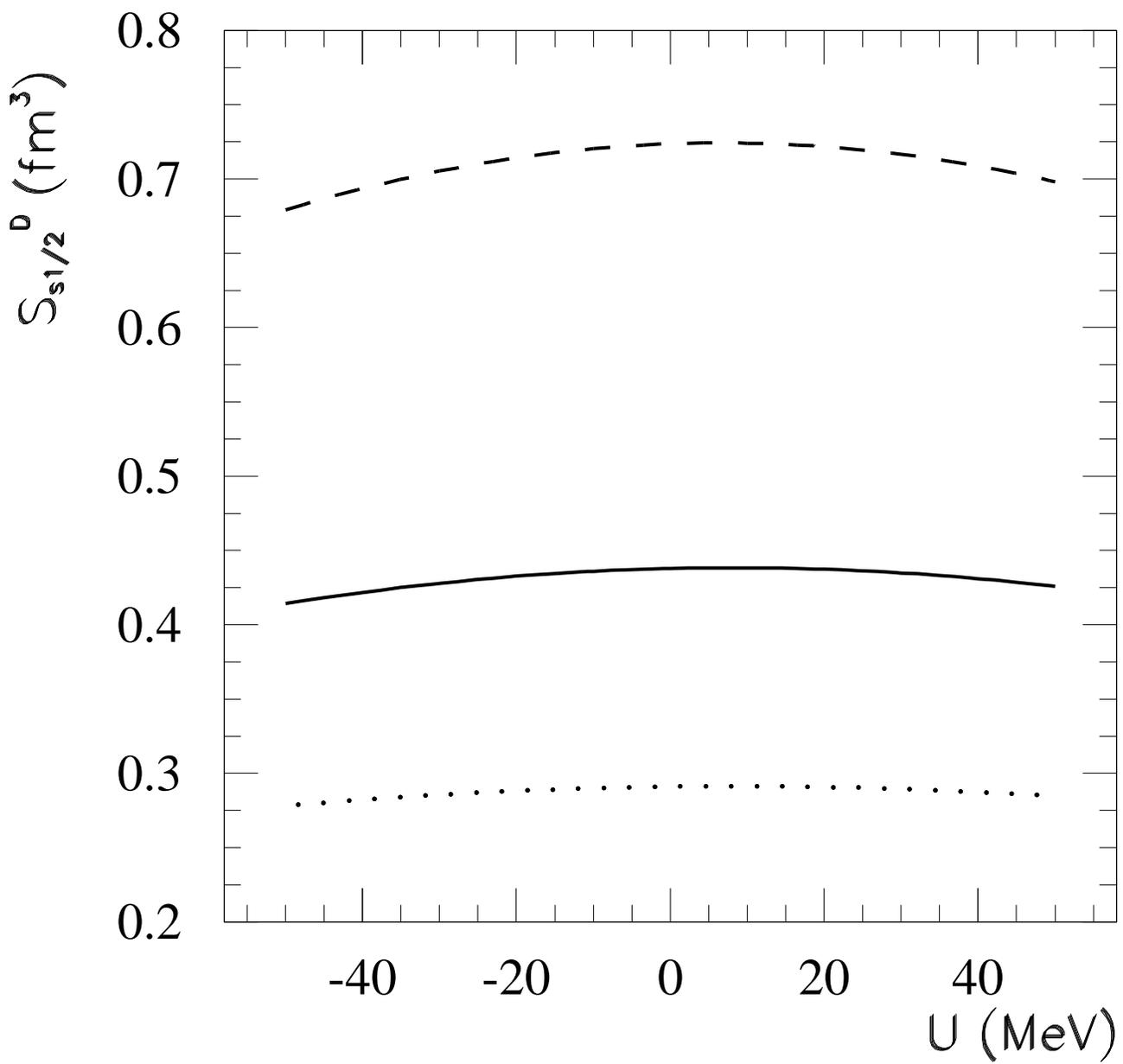
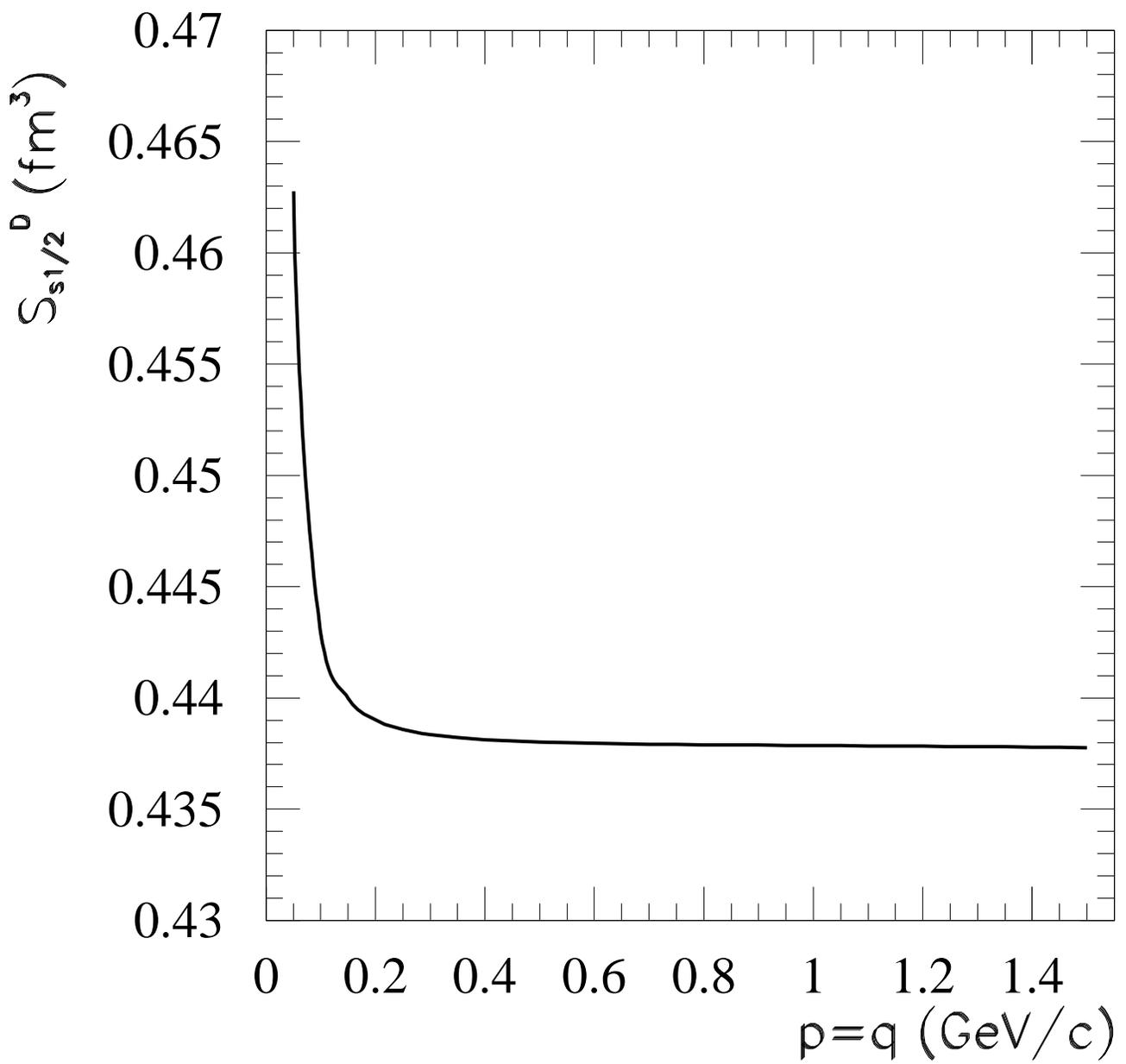
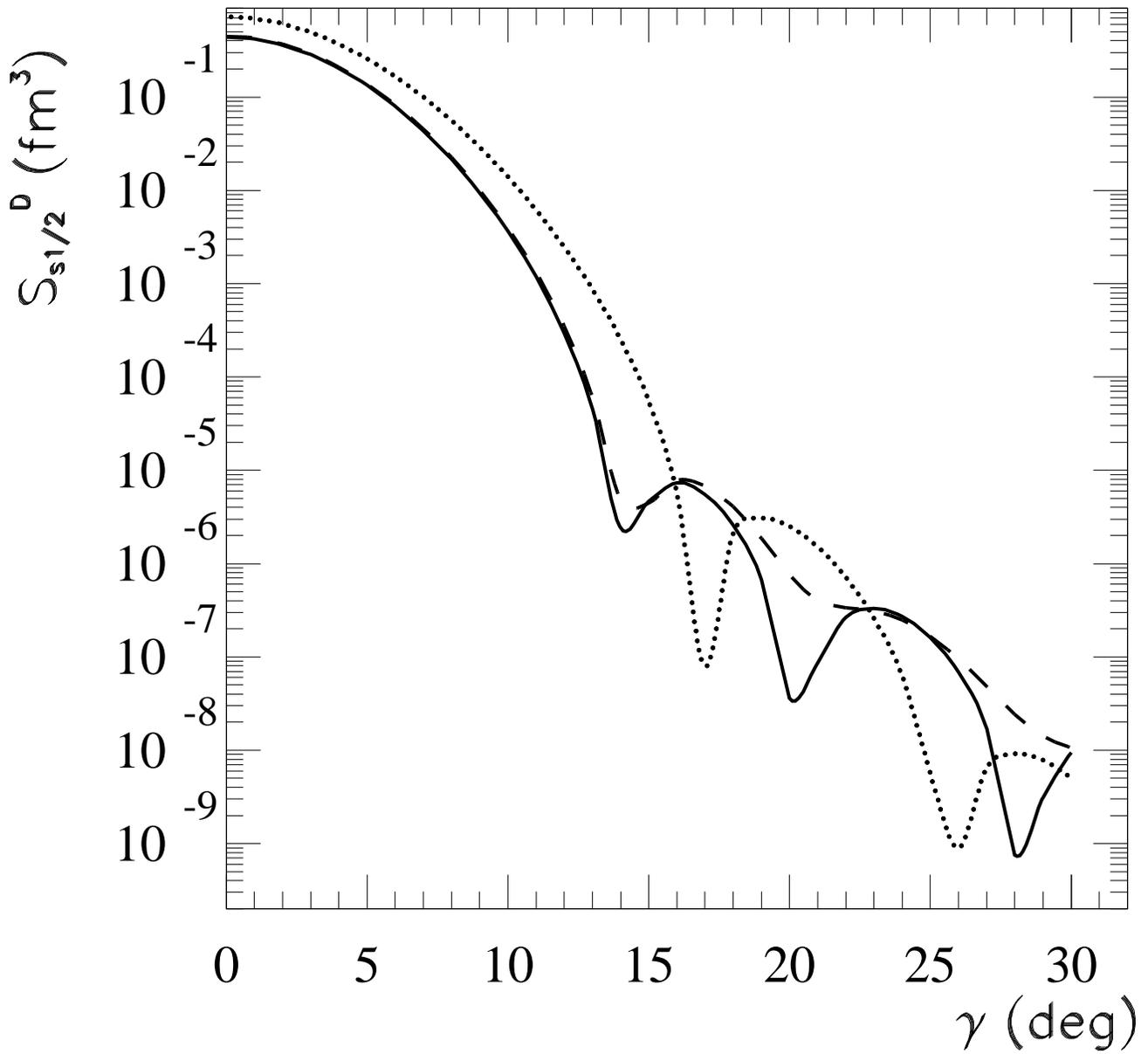
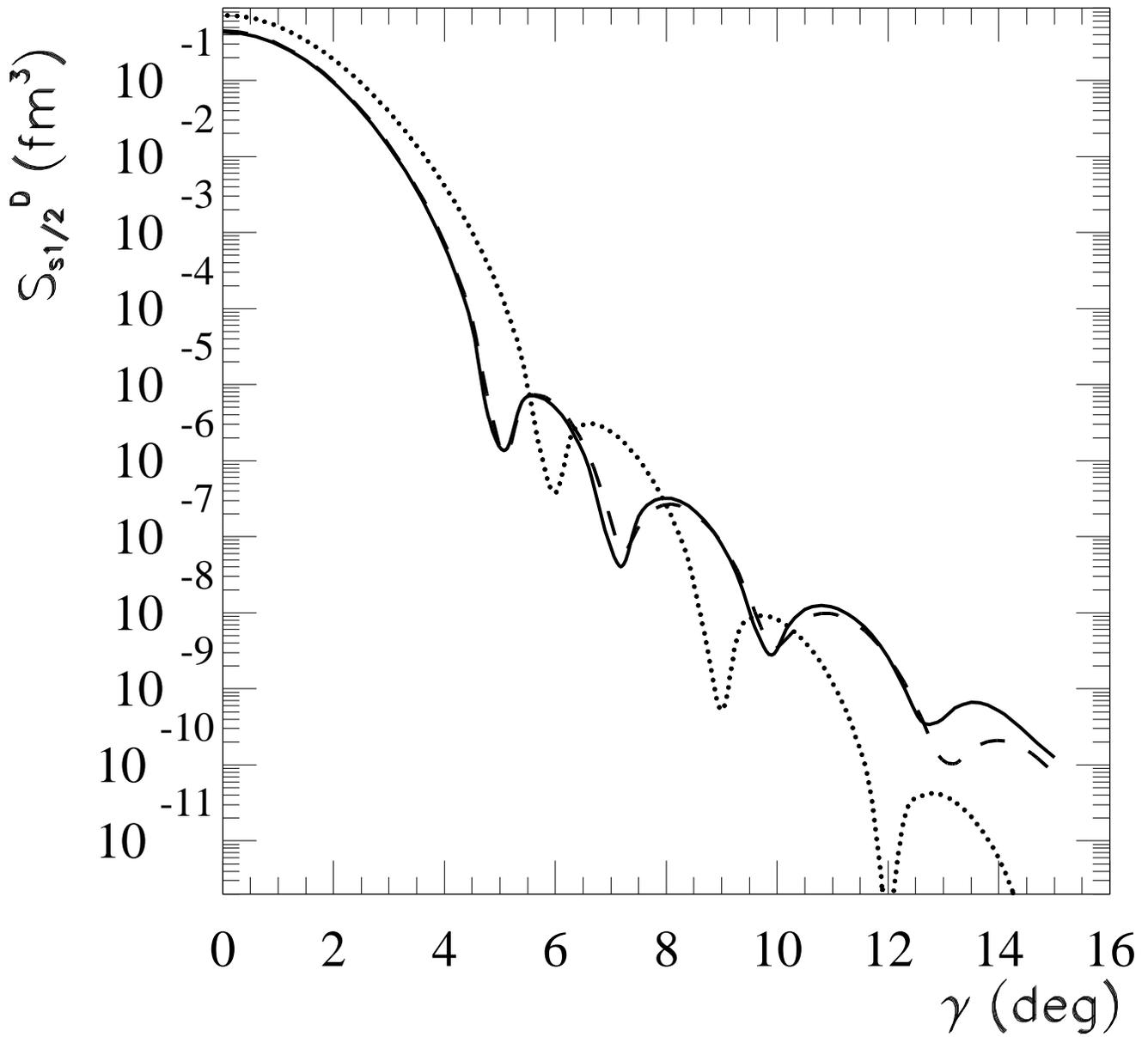
\end{document}